\documentclass{article}
\usepackage{spconf,amsmath,graphicx,float}
\usepackage{booktabs}
\usepackage{multirow}
\usepackage{hyperref}

\title{Importance of Different Temporal Modulations of Speech: \\
A Tale of Two Perspectives}
\name{Samik Sadhu$^1$, Hynek Hermansky$^{1,2}$}
\address{
  $^1$Center for Language and Speech Processing, Johns Hopkins University, USA\\
  $^2$Human Language Technology Center of Excellence, Johns Hopkins University, USA}

\begin{document}
\ninept
\maketitle
\begin{abstract}
  How important are different temporal speech modulations for speech recognition? We answer this question from two complementary perspectives. Firstly, we quantify the amount of phonetic \textit{information} in the modulation spectrum of speech by computing the mutual information between temporal modulations with frame-wise phoneme labels. Looking from another perspective, we ask - which speech modulations an Automatic Speech Recognition (ASR) system prefers for its operation. Data-driven weights are learned over the modulation spectrum and optimized for an end-to-end ASR task. Both methods unanimously agree that speech information is mostly contained in slow modulation. Maximum mutual information occurs around 3-6 Hz which also happens to be the range of modulations most preferred by the ASR. In addition, we show that the incorporation of this knowledge into ASRs significantly reduces their dependency on the amount of training data. 
\end{abstract}
\section{Introduction}
Modulation spectrum of speech has traditionally referred to the magnitude spectrum of its temporal envelope \cite{greenberg1996insights,hermansky1997modulation}. Distortion of these temporal dynamics adversely affects speech recognition. Generally, in modern Automatic Speech Recognition (ASR) systems, speech features are generated at a fixed sample rate of 100 Hz. On the front end side, speech dynamics are integrated into the ASR features by using delta and double-delta features or directly feeding raw or smoothed speech temporal patterns or TRAPS \cite{hermansky1998traps,hermansky1999temporal,athineos2004lp} over a long context into the ASR. 

However, the importance of different modulation frequencies for speech recognition is not immediately evident. Prior works have analyzed the effect of distorting different modulation frequencies on speech recognition  \cite{drullman1994effect,kanedera1997importance,arai1996intelligibility} and concluded that most important temporal modulations occur below 30 Hz - modulations around 4 Hz being the most important. Alternate studies analyzing the syllabic rates of speech \cite{greenberg1997origins} indicate that a majority of modulation energy lies within 2-8 Hz with a peak at 4 Hz. LDA-based derivation of temporal filters (RASTA filters) also show that slow modulations hold the most information for discriminating phonemes \cite{van1997data}. 

In this work, we look at two yet unexplored ways of investigating the modulation spectrum of speech. Although in some prior studies, the phase component of modulation has also been shown to bear speech information \cite{kanedera1998properties}, we primarily analyze the traditionally defined magnitude component of speech modulation, approaching it from two perspectives. On one hand, without performing speech recognition, we analyze phonetic information in speech, and on the other hand, we let a state-of-the-art machine decide which modulations to emphasize for speech recognition. 

\section{Computing Speech Modulations}
\label{sec:fdlp}
We use Frequency Domain Linear Prediction (FDLP) to approximate the power envelope of the speech signal in different frequency sub-bands over a long context of 1.5 seconds. The magnitude spectrum of these approximated log-transformed envelopes gives the modulation spectrum of speech in respective frequency sub-bands.

\subsection{Frequency Domain Linear Prediction (FDLP)}

At the heart of FDLP is the idea that linear prediction analysis of the inverse Fourier transform of a time domain signal gives an all-pole transfer function whose time response approximates the power of speech \cite{sadhu22_interspeech} - the level of approximation being controlled by the model order. 

\subsection{Computing Modulation Spectrum from FDLP}
The modulation spectrum can be obtained directly by computing the Fourier transform of the log FDLP response over 1.5 seconds of Hanning windowed speech. Computationally, this requires two Fourier transforms post linear prediction analysis. A time-saving alternative is to compute these modulation coefficients as the complex cepstrum of the FDLP model recursively from the linear prediction coefficients \cite{sadhu2019m}. The complex cepstrum of a ``complex" FDLP model is exactly equivalent to the modulation spectrum of speech \cite{sadhu22_interspeech}. 

\subsubsection{Spectral resolution of modulation spectrum}
FDLP analysis of a $T$ seconds long signal gives modulation coefficients at $1/T$ Hz resolution. In addition, the use of Hanning windows over $T$ seconds reduces the effective frequency resolution by half. This justifies our use of 1.5 second long context window to capture slow modulations in speech with sufficient spectral resolution.

\subsubsection{Modulation spectrum in different frequency sub-bands}
\begin{figure}[H]
    \centering
    \includegraphics[scale=0.30]{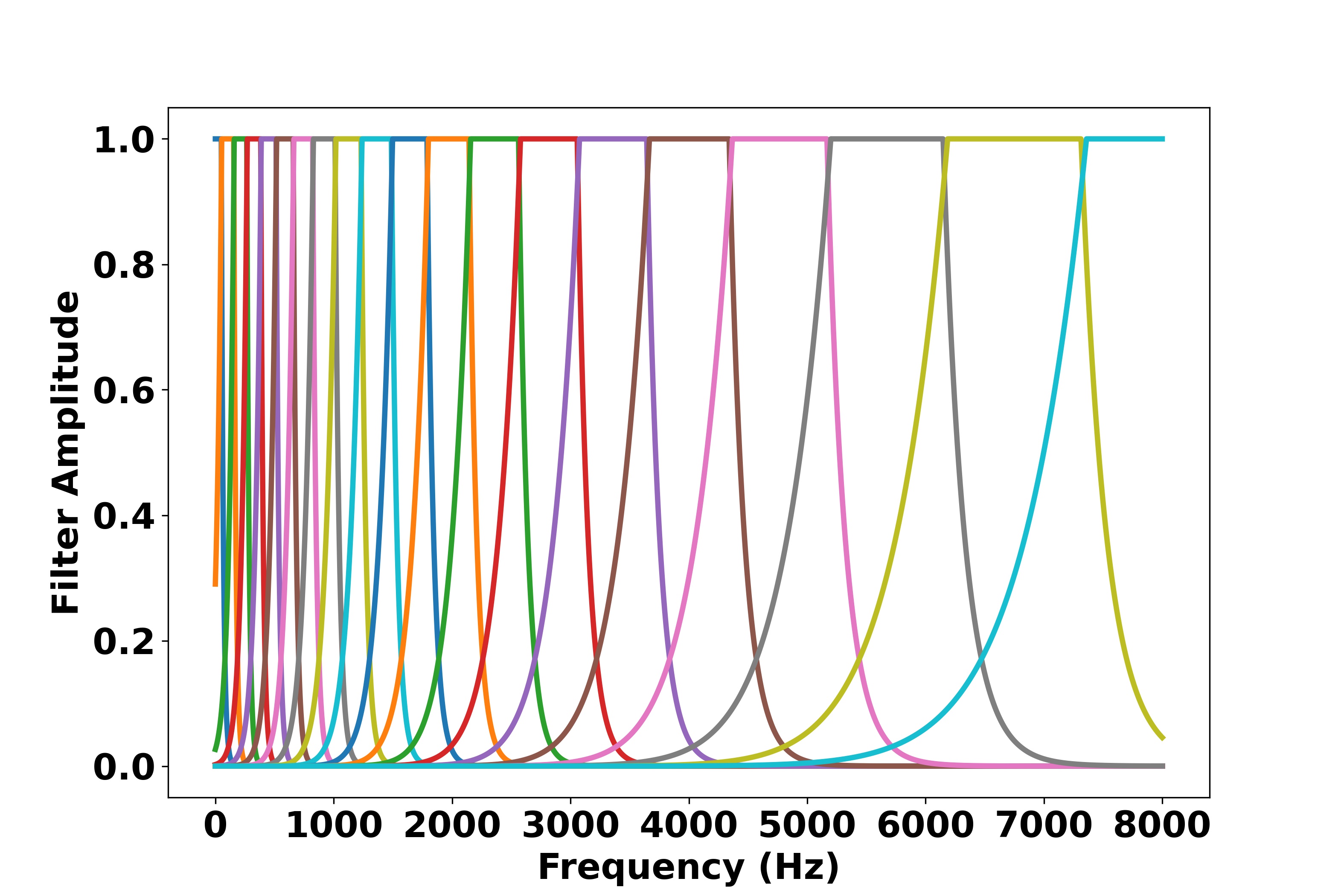}
    \caption{Modulation spectrum is computed in different cochlear filters}
    \label{fig:filter_bank}
\end{figure}

\begin{figure}[H]
    \centering
    \includegraphics[scale=0.33]{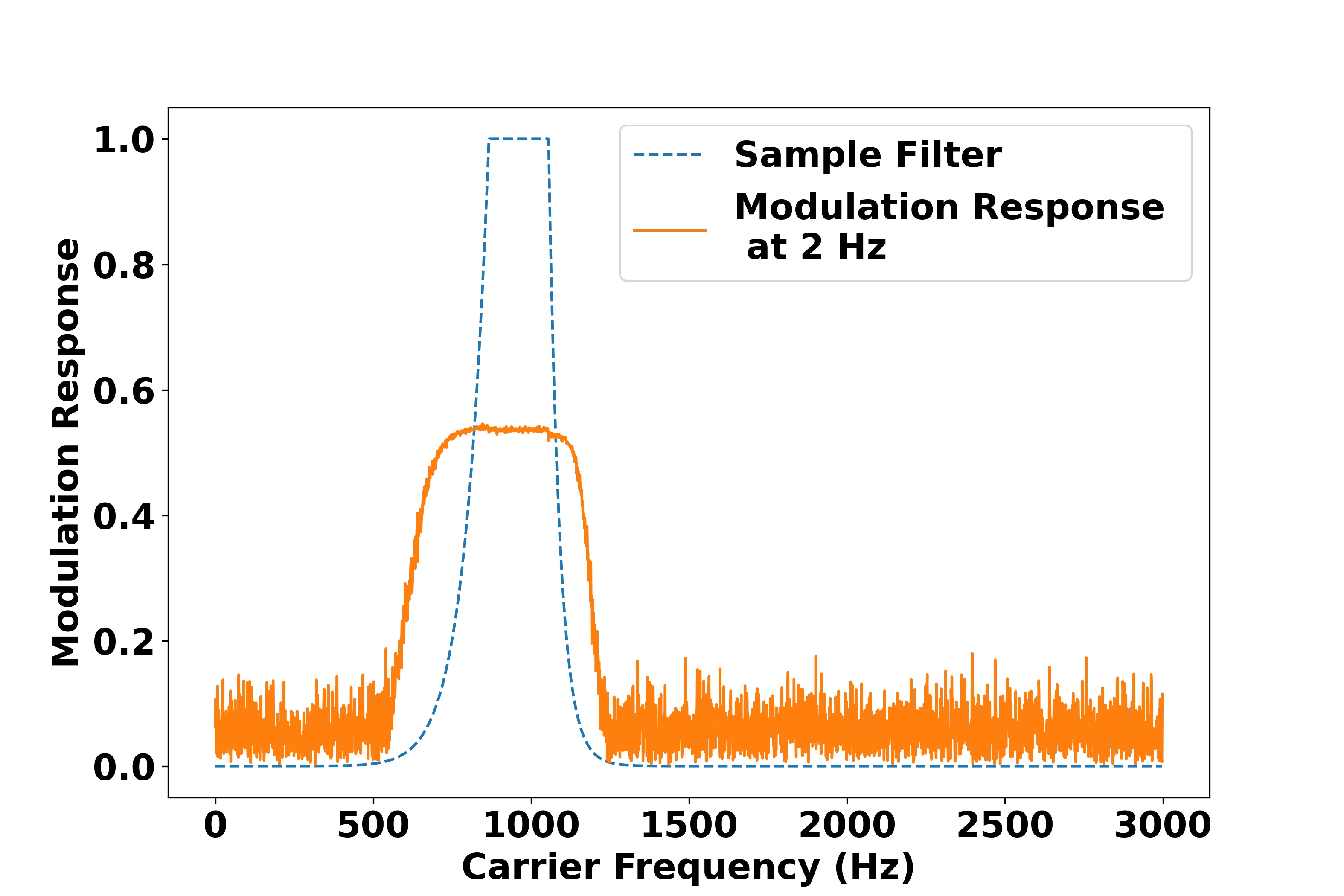}
    \caption{Amplitude modulated signals with modulation frequency 2 Hz and modulation depth 0.5 is generated with a variable carrier frequency. The inverse Fourier transform is weighted using the shown filter weight. The FDLP modulation spectrum at 2 Hz is shown as a function of the carrier frequency. The correct 2 Hz modulation magnitude of 0.5 is identified by our FDLP processing over the specified frequency range.}
    \label{fig:filter_response}
\end{figure}

To compute modulations in different frequency sub-bands, we \textit{weight} the inverse Fourier transform over appropriate frequency ranges. Linear Prediction analysis of this weighted inverse Fourier coefficients gives an all-pole model that approximates the power in the frequency sub-band specified by the weights. This allows modulations over any pre-defined range of carrier frequencies to be separated out (Figure \ref{fig:filter_response}). For our experiments, we design cochlear frequency band weights \cite{hermansky1990perceptual,sadhu2021radically} to analyze the input signal as shown in Figure \ref{fig:filter_bank}. 
\vspace{-20pt}

\section{Mutual Information Analysis of Modulation Spectrum}
\label{sec:MI}
Mutual Information (MI) between two random variables quantifies the information obtained about any one of the random variables by observing the other. Higher mutual information signifies greater mutual dependence between the two random variables. Mathematically, the MI between two discrete random variables $X$ and $Y$ is given by 
\begin{equation}
    I(X,Y)=\sum_{x \in \mathcal{X}}\sum_{y \in \mathcal{Y}}p_{X,Y}(x,y)\log\left( \frac{p_{(X,Y)}(x,y)}{p_{X}(x)p_{Y}(y)}\right)
\end{equation}
, where $P_X$ and $P_Y$ are the Probability Density Functions (PDF) of $X$ and $Y$, $P_{X,Y}$ is the joint PDF of $X$ and $Y$ and $X,Y$ takes values in the sample space $\mathcal{X} \times \mathcal{Y}$. 

\subsection{Generating Histograms of Modulations and Phoneme Labels}

Modulation spectrums are computed for each of 20 frequency sub-bands at a frame-rate of 100 Hz and multiplied with the modulation frequency to remove $1/f$ noise from the spectrum \cite{voss19751}. However, the modulation spectrum is a continuous-valued random vector. To empirically compute MI, we make a simplifying assumption that the modulation spectrum at different frequencies and across frequency sub-bands are statistically independent of each other. 
Continuous valued modulations at each modulation frequency and each filter is approximated by a discrete random variable $X$ following the same principles as \cite{yang2000search}. Histograms of these modulations are generated with 100 bins equally divided between the highest and lowest values obtained from all available data. 

We define frame-wise labels over 48 phoneme classes as the random variable $Y$. These labels are obtained by forced aligning the data with a HMM-DNN hybrid ASR model trained with Kaldi ASR toolkit \cite{povey2011kaldi}. 

\section{Modulations and Machines}
With the advent of End-to-end ASRs it has become possible to learn any pre-defined parameters involved in speech recognition by back-propagating the gradients computed from the ASR loss.

\subsection{Learning Modulation Weights}
\label{subsec:learn_mod_wts}
\begin{figure}[H]
    \centering
    \includegraphics[scale=0.12]{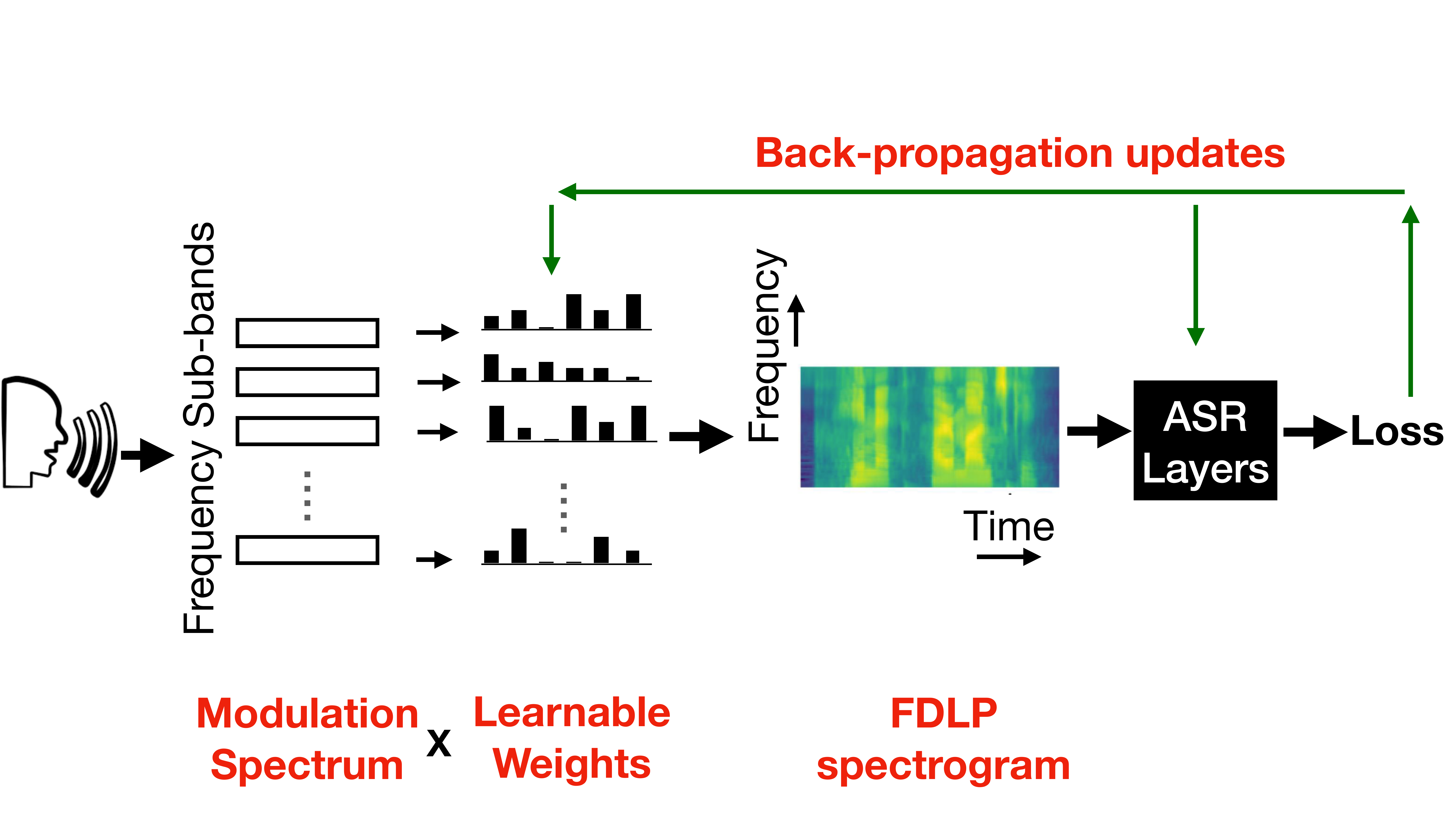}
    \caption{Modulations in different frequency sub-bands are multiplied with learnable weights that are updated with the ASR loss alongside the ASR parameters.}
    \label{fig:lifter_update}
\end{figure}

We modify a state-of-the-art ESPnet speech recognition model \cite{watanabe2018espnet} following a \texttt{espnet2} recipe with added  FDLP-spectrogram \cite{sadhu2021radically} front-end layers. In this fully end-to-end implementation, we compute modulation spectrum in different frequency sub-bands and \textit{weight} them by independently updated positive parameters in different sub-bands. The weighted modulations are transformed into FDLP-spectrograms that can be conveniently processed by the subsequent ASR layers (Figure \ref{fig:lifter_update}). The learned modulation weights are analyzed in section \ref{sec:results_modweights}.

\subsection{Knowledge is Recyclable}
Knowledge about the importance of different speech modulations learned from one training data in the form of modulation weights can be transferred to another ASR model that is to be trained on different training data. Already pre-trained modulation weights are kept fixed after being initialized into another ASR, the ASR layers themselves are only updated while training. Such prior knowledge is especially useful when enough training data is not available for ASR training.

\section{Experimental Details}

\subsection{Data sets}
We perform MI analysis on two different data sets. 
\begin{enumerate}
    \item \textbf{Wall Street Journal (WSJ):} This data consists of approximately 70 hours of read speech recorded in a clean studio environment. 
    \item  \textbf{REVERB \cite{kinoshita2013reverb}:} WSJ data is convolved with various room impulse responses to simulate reverberated data together with the addition of some background noise.
    \item \textbf{Librispeech \cite{7178964}}: We also use a 30 hours split of Librispeech for some additional experiments in section \ref{subsec:phase}.
\end{enumerate}

\subsection{FDLP-spectrogram configuration}
We compute FDLP-spectrograms with 1.5 second long Hanning windows, model order 80, and use 80 modulation coefficients that are computed at $\frac{1}{1.5}\approx 0.67$ Hz resolution.

\subsection{End-to-end ASR configuration}
 Data-driven modulation weights are derived by training the default transformer-based \cite{vaswani2017attention} ESPnet ASR (12 encoder layers with convolutional sub-sampling layers, 6 decoder layers and joint CTC-attention loss) with FDLP-spectrogram front-end with 15 hours of simulated reverberated speech from the REVERB data set. Alongside this, we use a transformer-based language model trained with text from the entire WSJ data. The modulation weights are all initialized to 1 over all 80 modulation coefficients. In our experiments, training the ASR parameters and the modulation weights together steered the model parameters into sub-optimal regions. To mitigate this problem, we trained the model for a total of 150 epochs, keeping the initialized modulation weights fixed for the first 60 epochs of training and eventually updating them together with the ASR parameters for the remainder of the epochs. 
 
 To study the knowledge transfer capability of these modulation weights, we train a less complex 6 transformer layered CTC ASR model with FDLP-spectrogram front-end on small amounts of the WSJ data that is sampled at random from the entire WSJ data set. In this case, the modulation weights are not initialized to 1, but instead with the modulation weights already learned on the REVERB data and kept fixed during ASR training.  

\section{Results}
\label{sec:results}
\subsection{Where does information lie in the modulation domain?}
\subsubsection{Results on WSJ}
\begin{figure}[h]
    \centering
    \includegraphics[scale=0.35]{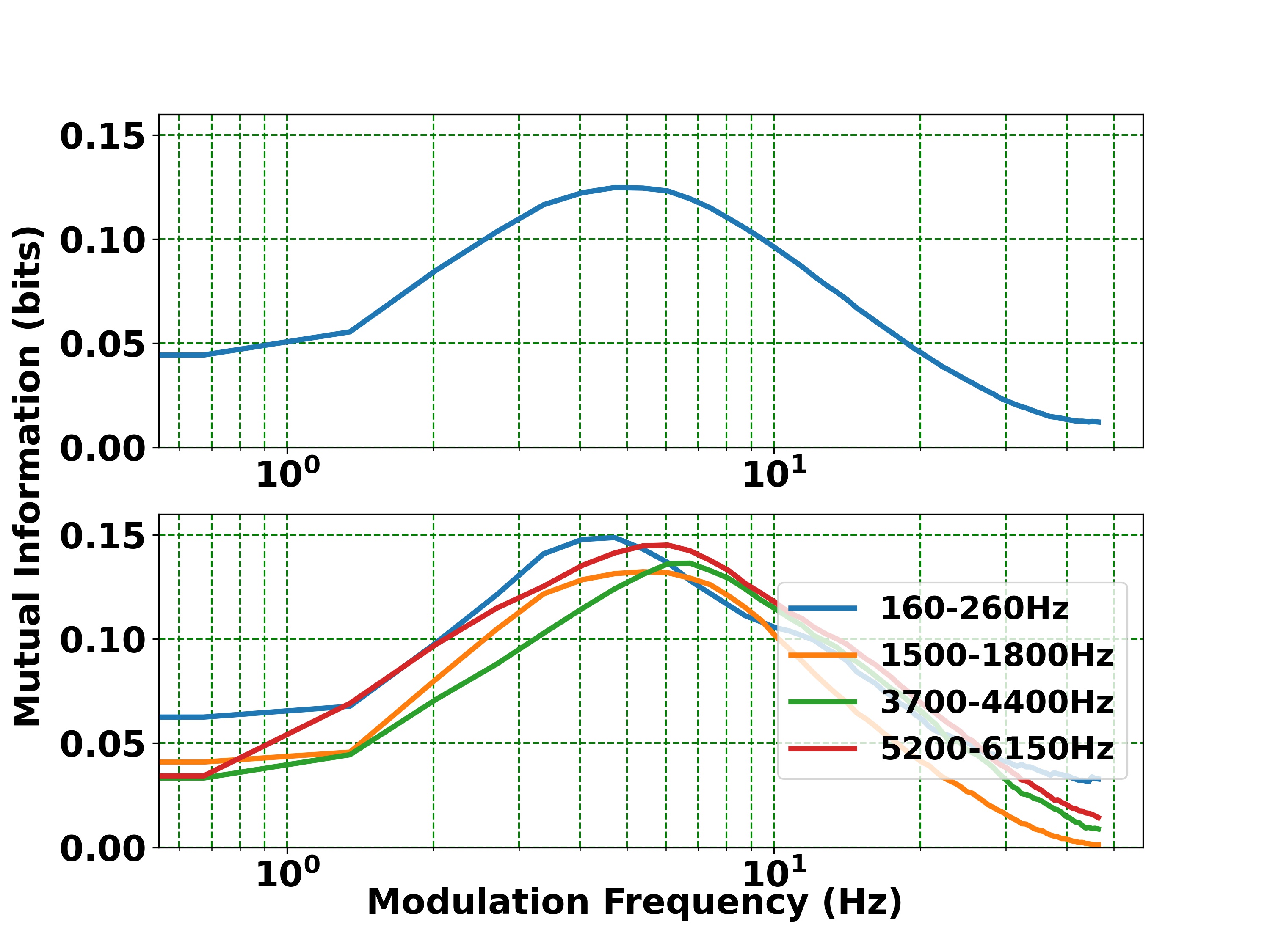}
    \caption{MI between modulation spectrum and phoneme labels for WSJ data \\
    (Top) Average MI from 20 frequency sub-bands \\
    (Bottom) MI in 4 filters spread across the whole frequency range}
    \label{fig:MI_1}
\end{figure}

Figure \ref{fig:MI_1} shows the average MI from all 20 filters as well as individual MIs in four different filters spread across the spectrum for clean read WSJ speech. From the average MI, we can see that most information lies below 40 Hz with maximum information concentrated around 3-6 Hz. These observations are consistent with previous studies on the importance of speech modulations using human and machine recognition of  speech \cite{drullman1994effect,kanedera1997importance,arai1996intelligibility}. Peak information at lower frequency sub-bands occurs at 4-5 Hz, but for the higher sub-bands the peak shifts to a higher range of 6-7 Hz. The peak energy in the modulation spectrum of speech also shows similar traits \cite{greenberg2004essential}.  

\begin{figure}[h]
    \centering
    \includegraphics[scale=0.35]{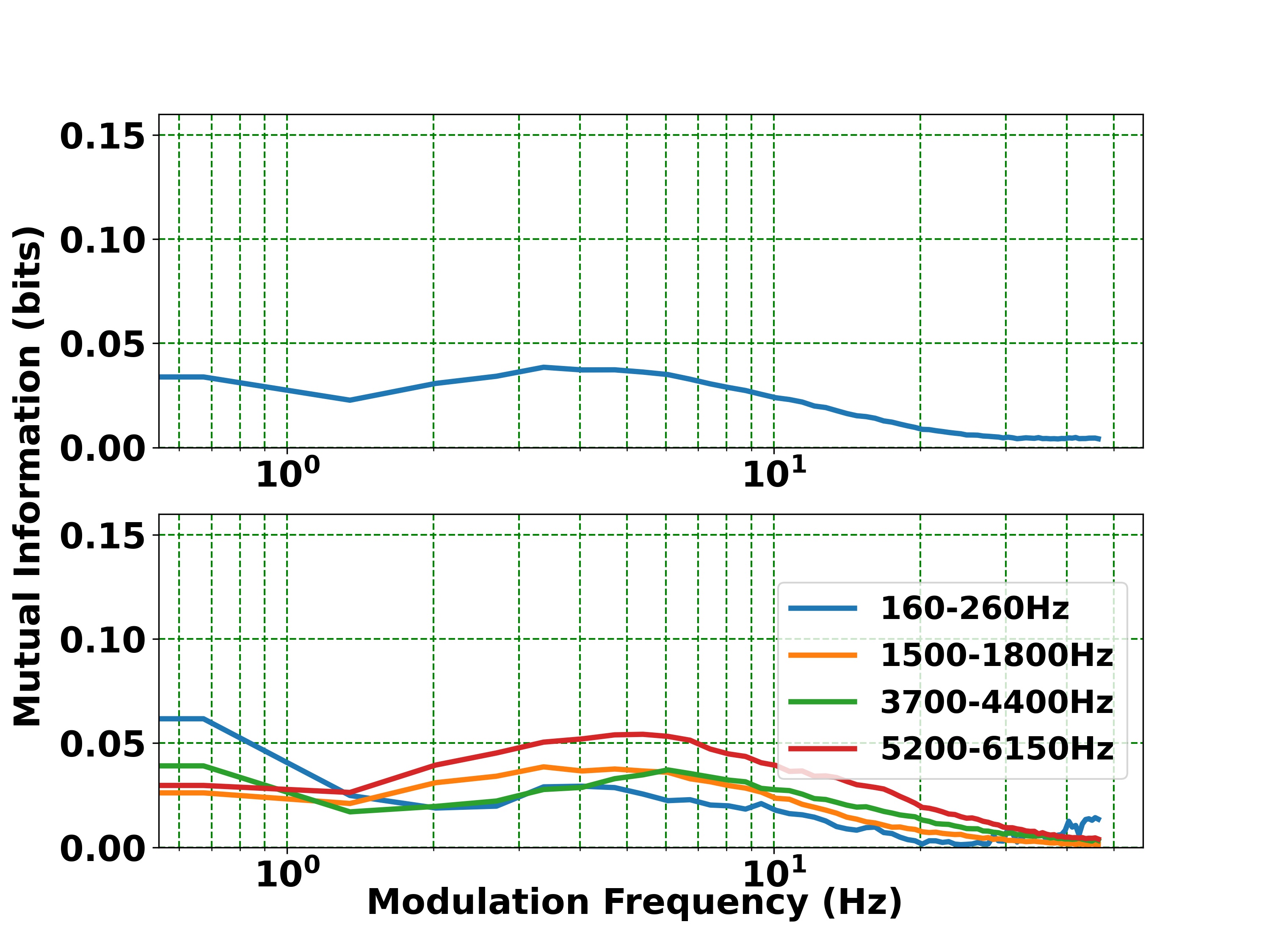}
    \caption{MI between modulation spectrum and phoneme labels for REVERB data \\
    (Top) Average MI from 20 frequency sub-bands \\
    (Bottom) MI in 4 filters spread across the whole frequency range}
    \label{fig:MI_2}
\end{figure}

\vspace{-20pt}
\subsubsection{Results on REVERB}
There is a significant reduction in MI for corrupted REVERB data with the information at 3-6 Hz getting reduced by $\approx \frac{1}{3}$ of clean read speech (Figure \ref{fig:MI_2}). This observation is not entirely surprising since machine speech recognition performance on REVERB with single channel reverberated speech data is significantly worse as compared to WSJ \cite{sadhu2021radically}.  Reverberations have also been shown to impact 4-6 Hz modulations in speech transmission and to drastically reduce human speech intelligibility  \cite{houtgast1985review,greenberg2004essential}.

\subsection{Which modulations do machines like?}
\label{sec:results_modweights}
\begin{figure}[H]
    \centering
    \includegraphics[scale=0.31]{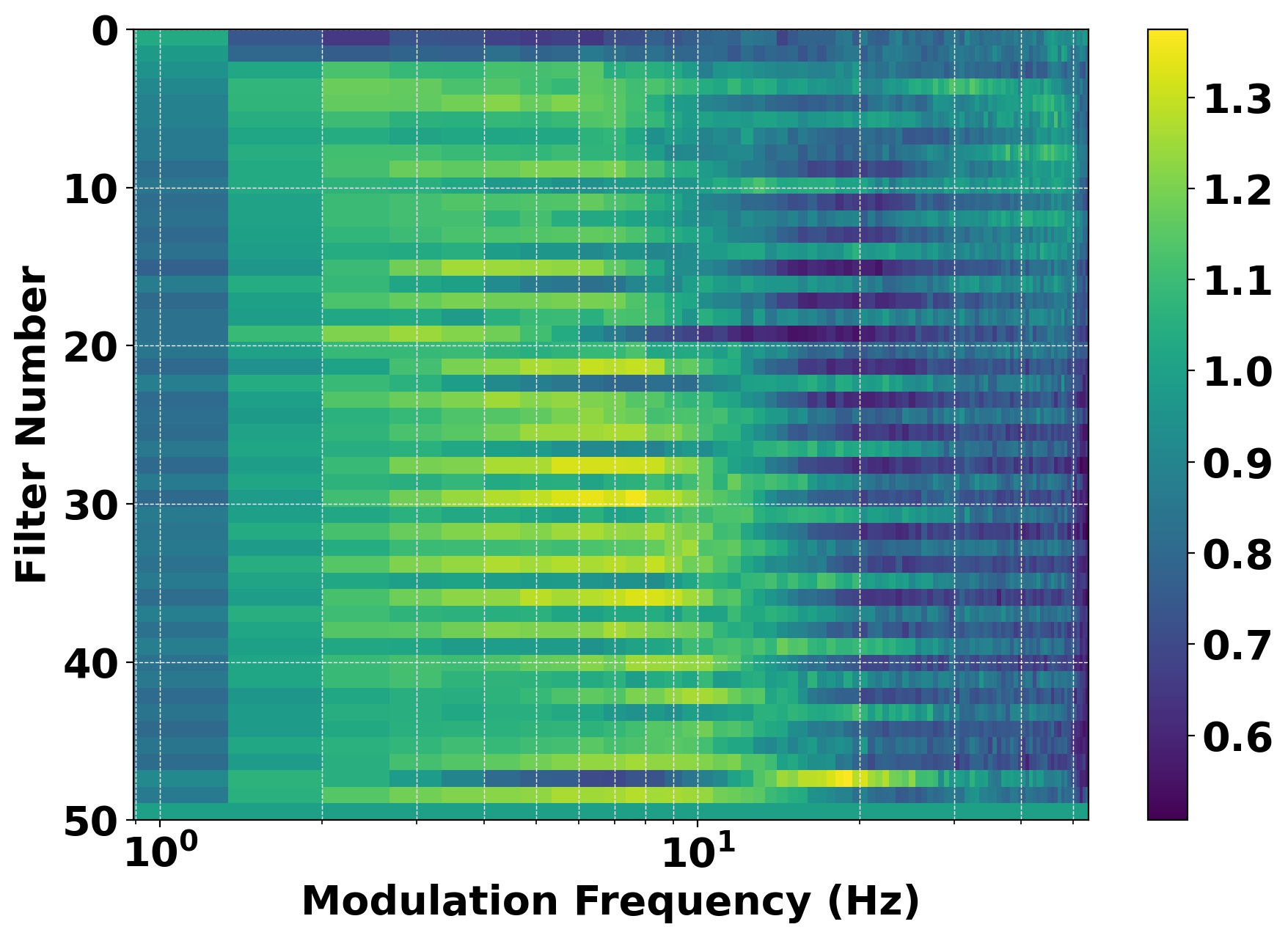}
    \caption{Modulation weights learned over 50 frequency sub-bands on 15 hours of REVERB speech data}
    \label{fig:lifter_allbanks}
\end{figure}

\begin{figure}[h]
    \centering
    \includegraphics[scale=0.31]{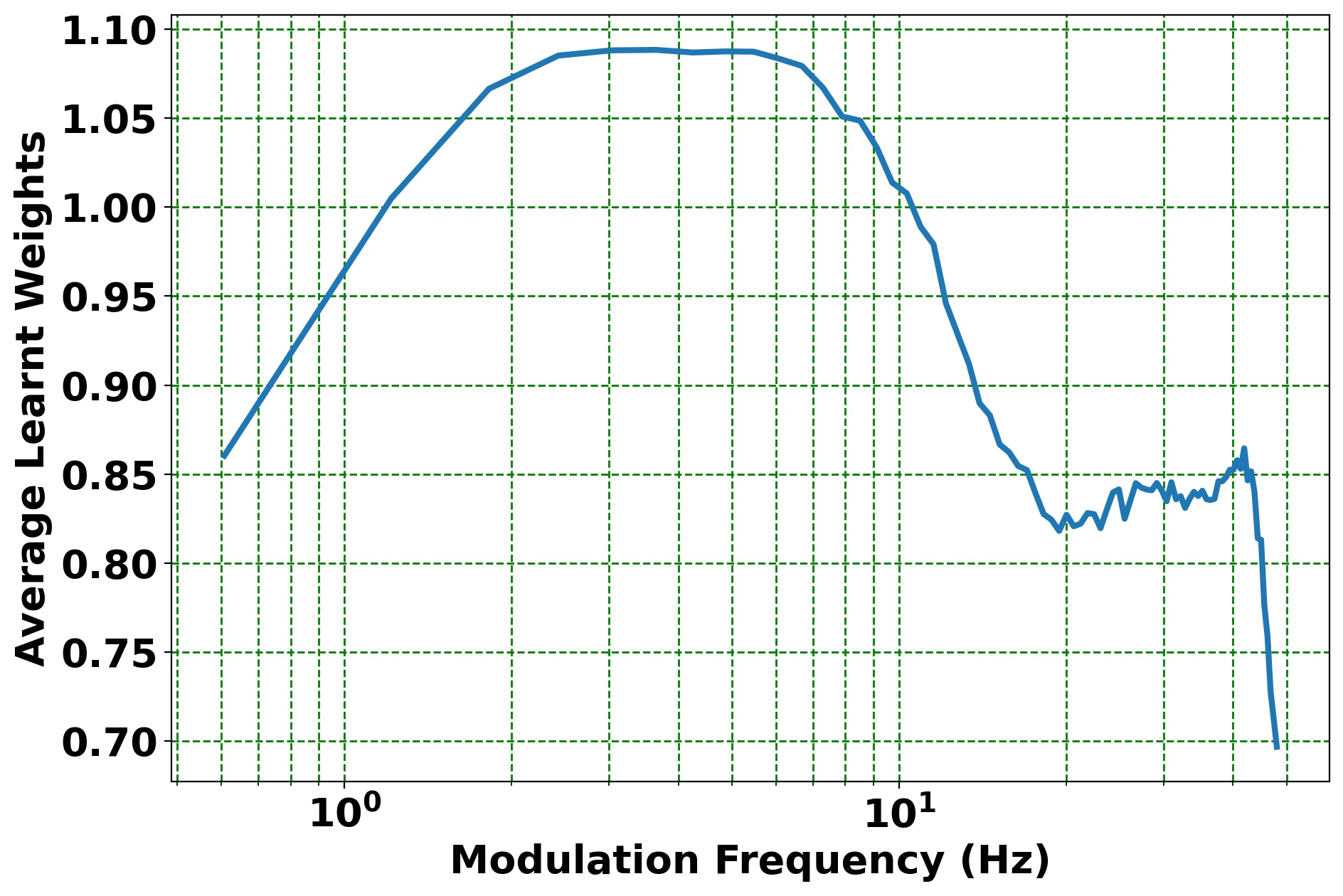}
    \caption{Average modulation weights learned over 50 frequency sub-bands on 15 hours of REVERB speech data}
    \label{fig:lifter_avg}
\end{figure}

Figure \ref{fig:lifter_allbanks} shows that the ASR assigns more weight over slow modulations in a vast majority of the frequency sub-bands. The average weights across all frequency bands shown in Figure \ref{fig:lifter_avg} show a striking similarity to hand-crafted RASTA filters \cite{hermansky1994rasta} as well as LDA-based design of RASTA-like filters \cite{van1997data}. Maximum weight is applied to 3-6 Hz modulations which we have also shown to contain maximum MI with phoneme labels.  

\subsection{Are the modulation weights useful?}
Table \ref{tab:results1} shows that modulation weights learned on 15 hours of REVERB data are indeed beneficial when used in a different ASR trained on less than 15 hours of WSJ data - the advantage diminishes when the WSJ training data is scaled up to 15 hours. Interestingly, the use of pre-trained modulation weights makes the ASR less sensitive to the amount of training data and the WER\% increases gracefully with a reduction in the amount of training data. On the other hand, there is a drastic jump in WER\% with a reduction in the amount of data when modulation weights are fixed to 1. In conclusion, the dependence of deep networks on large amounts of data can be reduced by making the ASR knowledgeable about which speech modulations to emphasize. 

In our experiments, updating the already pre-trained weights on the new data did not result in any significant gains.

\begin{table}[H]
\centering
\renewcommand{\arraystretch}{1.2}
\begin{tabular}{ccc}
\hline
\multirow{2}{*}{\begin{tabular}[c]{@{}c@{}}Hours of \\ data\end{tabular}} & \multicolumn{2}{c}{WER \%}                                                                          \\ \cline{2-3} 
                                                                          & Modulation wts. =1 & \begin{tabular}[c]{@{}c@{}}Modulation wts.\\  re-used from REVERB\end{tabular} \\ \hline
7                                                                         & 45.6               & 34.7                                                                           \\
9                                                                         & 50.7               & 29.3                                                                           \\
15                                                                        & 29.5               & 29.0                                                                           \\ \hline
\end{tabular}
\vspace{20pt}
\caption{Results from simple CTC ASR models trained on varying amounts of WSJ data are shown. The first column of results shows WER\% from a model trained with the modulation weights fixed at 1, whereas the second column shows the WER\% when, before training,  the modulation weights are replaced by those learned on the REVERB data. These weights are not updated over the training duration.}
\label{tab:results1}
\end{table}
\subsection{Effect of modulation phase}
\label{subsec:phase}
Although this paper focuses on the magnitude modulation spectrum, we present a preliminary study on the effects of learning weights on the complete modulation spectrum. In order to do so, two separate weights are defined, one on the real part and one on the imaginary part of the modulation spectrum, each updated independently with the ASR loss. The results are shown in Table \ref{tab:results2}.
\begin{table}[H]
\centering
\begin{tabular}{lc}
\hline
ASR type                       & WER \% \\ \hline
modulation weights fixed = 1               & 20.7   \\
magnitude weights learned          & 19.7   \\
real and imaginary weights learned & 19.4   \\ \hline
\end{tabular}
\caption{ASR WER\% are shown on 30 hours of Librispeech data. The first row shows the baseline ASR result with modulation weights fixed at 1. The second row shows the ASR results when weights on the magnitude of the modulation spectrum are learned as in section \ref{subsec:learn_mod_wts} while in the last row, independent weights on the real and imaginary parts of the modulation spectrum are learned.}
\label{tab:results2}
\end{table}
\vspace{-10pt}
These results, on one hand, show that learning magnitude weights on the modulation spectrum can improve ASR performance over fixed weights. On the other hand, when weights are learned on the entire modulation spectrum thereby modifying its magnitude and phase, there is further improvement in the WER\%. This initial observation is supported by prior studies on the importance of the modulation phase in speech recognition and needs further exploration in future work. 
\vspace{-10pt}
\section{Conclusions}
In this paper, we show that information theoretic analysis of raw speech data, as well as machine recognition of speech both, lead to the ultimate conclusion that slow speech modulations constitute the highest information-bearing components of the modulation spectrum of speech. 

Additionally, the use of this prior information for machine recognition of speech results in significant gains in ASR performance and makes the ASR less dependent on the amount of training data. 
\vspace{-10pt}
\section{Acknowledgements}
This work was supported  by the Center of Excellence in Human Language Technologies, The Johns Hopkins University, and by a gift from Google Research.

\vfill\pagebreak

\bibliographystyle{IEEEbib}
\bibliography{refs}

\end{document}